\def\bq{\begin{equation}}
\def\eq{\end{equation}}
\def\beq{\begin{equation}}
\def\bqa{\begin{eqnarray}}
\def\eqa{\end{eqnarray}}
\def\bqb{\begin{eqnarray*}}
\def\eqb{\end{eqnarray*}}
\def\to{\rightarrow}
\def\lsim{\raise0.3ex\hbox{$<$\kern-0.75em\raise-1.1ex\hbox{$\sim$}}}
\def\gsim{\raise0.3ex\hbox{$>$\kern-0.75em\raise-1.1ex\hbox{$\sim$}}}

\def\llgm{\left\lgroup\matrix}
\def\rrgm{\right\rgroup}
\def\lsim{\raise0.3ex\hbox{$<$\kern-0.75em\raise-1.1ex\hbox{$\sim$}}}
\documentstyle[12pt,epsf]{article}
\renewcommand{\theequation}{\arabic{section}.\arabic{equation}}
\global\nulldelimiterspace = 0pt

\def\roughly#1{\mathrel{\raise.3ex
    \hbox{$#1$\kern-.75em\lower1ex\hbox{$\sim$}}}}
\def\lsim{\roughly<}
\def\gsim{\roughly>}
 
\begin{document}
\pagenumbering{arabic}
\thispagestyle{empty}
\def\thefootnote{\fnsymbol{footnote}}
\setcounter{footnote}{1}
 
\begin{flushright}
BI-TP 99/30 \end{flushright} 
\vspace {4 cm} 
\begin{center} {\Large \bf HIGH-ENERGY CROSS SECTION FOR}\\[6mm] 
{\Large \boldmath{$e^+e^- \to W^+W^- \to {\rm \bf 4~ FERMIONS} 
(+ \gamma)$}\footnote{Supported by the Bundesministerium f\"ur Bildung
und Forschung, No.05 HT9PBA 2, Bonn, Germany.}}

\vspace{10mm}
{\large Y. Kurihara$^a$, M. Kuroda$^b$ and D. Schildknecht$^c$}\\[1.1ex]
\vspace{0.2cm}
$^a$ KEK, Tsukuba, Japan\\ 
\vspace{0.5cm}
$^b$ Institute of Physics, Meiji-Gakuin University\\
Yokohama, Japan.\\
\vspace{0.5cm}
$^c$ Fakult{\"a}t f{\"u}r Physik, Universit\"at Bielefeld\\
Bielefeld, Germany\\
\vspace{1cm}
\end{center}
\vspace{ 2 cm}
\leftline{\Large \bf Abstract}

A recently suggested high-energy Born-form representation of the one-loop
helicity amplitudes for $e^+e^- \to W^+W^-$ is supplemented by including
$W^\pm$ decay and hard-photon radiation. Results for the differential and
the total cross section for 
$e^+ e^- \to W^+ W^- \to 4 {\rm fermions} (+ \gamma)$
are given for the high-energy region of $\sqrt s \gsim 500 GeV$.

\vspace {2 cm}

\begin{flushleft}
August 1999
\end{flushleft}

\vfill\eject

\section{Introduction}

In a recent paper on W-pair production in $e^+e^-$ annihilation by two of
us \cite{Kuschi} a high-energy representation for the one-loop-corrected
helicity amplitudes was given in simple analytic form. The one-loop
helicity amplitudes in Ref. \cite{Kuschi} are represented in the form
of a Born approximation, the weak and the electromagnetic coupling
constant of the Born approximation being replaced by (three)
invariant amplitudes that depend on the kinematic variables of the reaction,
the energy, momentum transfer and soft-photon cut-off. The
representation of the helicity amplitudes differs from previous 
ones \cite{Didi} with respect to the choice of the covariants and the
corresponding invariant amplitudes. In distinction from Ref. \cite{Didi},
the explicit and fairly simple analytic form of the invariant amplitudes,
written down in a few lines, is a novel feature of Ref. \cite{Kuschi}.
In the derivation of the expressions for the invariant amplitudes in 
Ref. \cite{Kuschi}, the high-energy expansions of the cross sections
of Ref. \cite{Been} were extensively used.\par

     In the present work we will supplement our recent results by including
the decay of the produced  $W^\pm$ bosons as well as hard-photon
radiation.

\section{The ansatz for $\bf e^+e^- \to W^+W^- \to 4 {\rm\bf fermions}$, 
generalities}

For a theoretical description of W-pair production including W decay,
as a natural starting point, we will use the (one-loop-corrected) 
amplitudes for on-mass-shell production and decay. The propagation of the
decaying W bosons will be taken care of by Breit-Wigner denominators with
constant widths. In so far as on-shell-production and -decay amplitudes
are employed, while the invariant masses of the fermion pairs only enter
via the Breit-Wigner denominators and the four-fermion phase space, such an
ansatz corresponds to a narrow-width approximation for $e^+ e^- \to W^+W^-
\to 4 {\rm fermions}$.  As on-shell
production and decay amplitudes are gauge invariant, the 
so-defined amplitudes for $e^+e^- \to W^+ W^- \to 4 {\rm fermions}$
are gauge-independent by construction. The approach should be particularly
reliable for invariant masses of the produced fermion pairs in the
vicinity of the $W^\pm$ mass, $M_W$.

Four-fermion production, $e^+ e^- \to 4 {\rm fermions}$, in general,
does not only proceed via the production and subsequent decay of a
W-boson pair. Additional contributions are present that involve only a
single W pole, or no pole at all (cf. e.g. \cite{Berends}). 
With respect to
W-pair production, such contributions form a non-resonant (more precisely,
a non-doubly-resonant) background. It is expected that by applying suitable
experimental cuts on the invariant masses of the produced fermion pairs,
one may be able to eliminate such background contributions to a large
extent when comparing theoretical predictions with experimental data.

  From the point of view of four-fermion production the above ansatz, based
on on-mass-shell amplitudes and $W^\pm$ Breit-Wigner denominators,
corresponds to a double-pole approximation (DPA) \cite{Ditt,Ditt1,W.Been}. 
The gauge invariance of the full (one-loop) four-fermion-production
amplitude implies that the double-pole residue in this amplitude is
gauge-parameter independent. In addition to the gauge-parameter-independent
on-mass-shell-production and -decay amplitudes, the double-pole residue
also contains so-called non-factorizable corrections \cite{Ditt},
essentially due to soft-photon exchange connecting, e.g. a decay fermion
from $W^+$- decay with the $W^-$-boson. Extensive investigations 
\cite{Denner,Fadin} led to the result that such corrections vanish, when
integrated over the invariant masses of the fermion pairs. Moreover, they
are entirely negligible \cite{Denner} at $e^+ e^-$ energies above a
few hundred GeV, the region of energies considered in the present
paper. Accordingly, non-factorizable corrections need not be considered
any further.

A quantitative analysis of the error induced by neglecting non-doubly-resonant
contributions to four-fermion production faces the difficulties of 
formulating a satisfactory gauge-invariant ansatz for
$e^+ e^- \to 4 {\rm fermions}$ that includes all (doubly-resonant, 
singly-resonant and
non-resonant) contributions. The simple replacement of $[k^2_\pm - M^2_W]^{-1}$
by $[k^2_\pm - M^2_W + i M_W \Gamma_W]^{-1}$ in general violates gauge
invariance as different parts in the gauge-invariant amplitude for
$e^+e^- \to 4 {\rm fermions}$ are differently affected by such a
replacement \cite{Berends}.
Nevertheless, it seems that such a ``fixed-width scheme'' finds some
justification in a ``complex-mass scheme'' \cite{Ditt1}. For an estimate
of the error induced by neglecting non-doubly-resonant contributions, a
fixed width ansatz should be quite reliable. Accordingly, we will use
it in section 3 to estimate the accuracy to be expected for the double-pole
approximation of four-fermion production.\par
     We note that how to go from the off-shell kinematics to the
on-shell kinematics is not unique.  In our analysis, in defining
the double-pole residue, we fix the production solid angles of 
$W^-$ and the decay solid angles of two of the
final fermions ( originating from $W^+$ and $W^-$) in the laboratory 
frame.\par

\section{Estimating Background Contributions}

With respect to W-pair production, four-fermion production not enhanced by
two intermediate W resonances, as mentioned in Section 2, 
 may be considered as a background. This
background, in general, contains four-fermion production via only one
intermediate W boson or via no W-resonance enhancements at all
\cite{Berends}.

In this section, we estimate the importance of such non-doubly-resonant
background contributions by comparing the results of tree-level
calculations for $e^+ e^- \to W^+W^- \to 4 {\rm fermions}$ 
with the results for $e^+e^- \to 4 {\rm
fermions}$ based on 
the full
set of diagrams contributing to the reaction; wherever W poles appear
in the full set of diagrams, as mentioned, fixed widths are introduced in
the denominators ("fixed width scheme"). Specifically, we will
consider the $u \overline{d} \overline{c} s$ final state, thus comparing
$e^+e^- \to W^+ (\to u \overline{d}) W^- (\to \overline{c} s)$ with 
$e^+ e^- \to u \overline{d} \overline{c} s$.  We will also
briefly comment on the semileptonic decay, 
$e^+e^-\to W^+(\to u\bar d) W^-(\to e \bar\nu_e)$.

Our numerical results are collected in Table 1. They are based on the
input parameters $M_W = 80.22 GeV, M_Z = 91.187 GeV$ and 
$\alpha = 128.07^{-1}$,
and were obtained by employing {\tt GRACE} \cite{Grace}, the automatic 
computation system for electroweak processes.  For simplicity,
the fermion masses are neglected in the calculation.  
The explicit calculation by {\tt GRACE} shows that the effect of the 
finite fermion masses is about 0.2\%.

\begin{table}
\begin{tabular}{|r|l|c|c|c|}\hline
Line & $\sqrt s$ & 500 GeV & 1 TeV & 2 TeV\\ \hline\hline
 1&$\sigma (e^+ e^- \to W^+W^-)$ & 7.458 & 2.785 & $9.421 \times 10^{-1}$\\ \hline\hline
\multicolumn{5}{|c|}{Zero width approximation}\\ \hline
2&$\sigma \times BR (W^+ \to u \overline{d}) \times BR (W^- \to \overline{c}s)$
&$8.289 \times 10^{-1}$ & $3.094\times 10^{-1}$ &
 $1.047 \times 10^{-1}$ \\ \hline\hline
\multicolumn{5}{|c|}{Breit-Wigner, full four-fermion phase space} \\ \hline
3&$\sigma (e^+ e^- \to W^+ (\to u \overline{d}) W^- (\to \overline{c} s))$
&$8.291 \times 10^{-1}$ & $3.097 \times 10^{-1}$ &
 $1.046 \times 10^{-1}$ \\
4&$\sigma (e^+ e^- \to u \overline{d} \overline{c} s)$ &
$8.466 \times 10^{-1}$ & $3.248 \times 10^{-1}$ &
 $1.124 \times 10^{-1}$ \\
5&Difference $\Delta$ in \% & 2.1 \% & 4.9\% & 7.5\% \\ \hline\hline
 \multicolumn{5}{|c|}{Breit-Wigner, restricted phase space, 
$\vert \sqrt{k^2_\pm} - M_W \vert \lsim 5 \Gamma_W$} \\ \hline
6&$\sigma (e^+ e^- \to W^+ (\to u \overline{d}) W^- (\to \overline{c} s))$
&$7.264 \times 10^{-1}$ & $2.713 \times 10^{-1}$ &
$9.16 \times 10^{-2}$ \\
7&$\sigma (e^+ e^- \to u \overline{d} \overline{c} s)$ & 
$7.275 \times 10^{-1}$ & $2.717 \times 10^{-1}$  & 
$9.19 \times 10^{-2}$ \\
8&Difference $\Delta$ in \% & 0.1 \% & 0.1 \% & 0.3 \% \\ \hline
\end{tabular}
\caption{Tree-level results in [pb] for $W^+W^-$-mediated four-fermion
production (specifically for the $u \overline{d} \overline{c} s$ final
state) compared with four-fermion production including (non-doubly-resonant)
background for different phase-space cuts.}
\end{table}

Line 1 of Table 1 gives the cross sections for (stable) W-pair
production for various energies. The results in line 2 of Table 1 are
 obtained
by multiplication of the W-pair-production cross sections by the appropriate
branching ratios, BR. The  branching ratios for $W^+ \to u \bar d$ 
and  $W^- \to \bar c s$ are 
given as $BR = 1/3$. Line 2 thus corresponds to the  zero-width 
approximation of $\Gamma_W \to 0, {\rm BR = const}$, i.e.,
$$
{{M_W \Gamma_W} \over {\vert k^2_\pm - M^2_W + i M_W \Gamma_W \vert^2}}
BR \to \pi \delta (k^2_\pm - M^2_W)\cdot BR.\eqno(3.1)
$$
The  results of line 2 practically agree with  the
results of line 3 that are obtained by (sevenfold Monte Carlo)
integration of the Breit-Wigner resonance shape over the 
four-fermion phase space using the on-shell W-pair-production
amplitudes.  The results in line 4 are based on
the full set of four-fermion-production diagrams with fixed
widths inserted where appropriate.  
The difference between the results in line 3 
and the four-fermion production, $e^+e^- \to u\bar d \bar c s$, of
line 4, depending on the energy,  lies between $2\%$ and $8\%$, 
as indicated in line 5.
Clearly, the entire four-fermion production exceeds 
the production via the decay of both $W^+$ and $W^-$
by $2\%$ to $8\%$.\par
     
     Finally, we enhance the relative contribution of the $W^+$ and $W^-$
resonances by imposing the restriction
$$
\vert \sqrt{k^2_\pm} - M_W \vert \lsim 5 \Gamma_W,\eqno(3.2)
$$
on the masses of the fermion pairs, $\sqrt{k_\pm^2}$, 
when integrating over the four-fermion phase space.  The results in lines 6
to 8 show that background contributions are reduced to roughly
0.1\% to 0.3\%, if the cut (3.2) is imposed.  Upon employing the cut (3.2), 
one can forget about all the four-fermion states that do not 
originate from the decay of two $W$-bosons.\par

     We add a comment at this point on the background and its
reduction by the cut (3.2), if the $W$ bosons are identified
by a decay mode different from the one discussed above.  In particular,
we have looked at the leptonic mode, $W^-\to e\bar\nu_e$.
In this decay mode, without cut,
the difference $\Delta$ between
$\sigma(e^+e^-\to W^+(u\bar d)W^-(e\bar\nu_e))$ and
$\sigma(e^+e^-\to u\bar d e \bar\nu_e)$, the quantity
corresponding to line 5 of Table 1, becomes very large,
81.9\%, 96.3\% and
99.3\%, for $\sqrt s=$ 500 GeV, 1 TeV and 2 TeV, respectively.
When the cut (3.2) on $k^2_\pm$ is applied, however, we can 
reduce the background
almost completely and the difference $\Delta$ corresponing
to line 8 of Table 1 becomes, 0.9\%, 0.4\% and 0.4\% respectively.\par

     Accordingly, it seems appropriate to apply theoretical as well as
experimental cuts of the kind (3.2), when comparing theoretical predictions
with experiment. This procedure largely circumvents the
gauge-invariance issues inherently connected with a general theoretical
treatment of four-fermion production.  
The procedure is conceptually simple and fully sufficient, moreover, 
for all practical purposes, as it is mainly the
empirical test of the W-boson properties that one is interested in.\par

\section{The ansatz for $\bf e^+ e^- \to W^+ W^- \to 4 {\rm\bf fermions}$,
 details.}

In one-loop order the helicity amplitudes of the process
$e^+(\sigma_+)e^-(\sigma_-) \to 
W^+_{\lambda_+}(f_1(\tau_1) \bar f_2(\bar\tau_2))$ 
$W^-_{\lambda_-}(f_4(\tau_4) \bar f_3(\bar\tau_3))$ are given by
\bq
     {\cal H}(\sigma_+,\sigma_-;\tau_1,\bar\tau_2;\tau_4,\bar\tau_3)
    = \sum_{\lambda_+,\lambda_-} 
          {{ {\cal H }(e^+(\sigma_+)e^-(\sigma_-) \to 
                   W^+_{\lambda_+} (f_1(\tau_1)\bar f_2(\bar\tau_2))
                   W^-_{\lambda_-} (f_4(\tau_4)\bar f_3(\bar\tau_3)))} 
              \over{K_+K_-}}, \label{4.1}
\eq     
where $\sigma_\pm$ is the positron(electron) helicity, $\lambda_\pm$
is the helicity of $W^\pm$, and $\tau_i (\bar\tau_i)$ are helicities
of the final fermions (antifermions).   The explicit form 
of the numerator of (\ref{4.1}) is given by
\bqa     
  &&  {\cal H }(e^+(\sigma_+)e^-(\sigma_-) \to
     W^+_{\lambda_+}(f_1(\tau_1)\bar f_2(\bar\tau_2))
     W^-_{\lambda_-}(f_4(\tau_4)\bar f_3(\bar\tau_3))) \label{4.2}  \\
  ~~~~~~&&=+{\cal H}_{Born}^{e^+e^-\to W^+W^-}(\sigma_+,\sigma_-;\lambda_+,\lambda_-)
      {\cal H}_{Born}^{W^+\to f_1\bar f_2}(\lambda_+;\tau_1,\bar\tau_2)
      {\cal H}_{Born}^{W^-\to f_3\bar f_4}(\lambda_-;\tau_4,\bar\tau_3)
      \nonumber \\
  &&~~~ +\delta{\cal H}^{e^+e^-\to W^+W^-}(\sigma_+,\sigma_-;\lambda_+,\lambda_-)
     {\cal H}_{Born}^{W^+\to f_1\bar f_2}(\lambda_+;\tau_1,\bar\tau_2)
     {\cal H}_{Born}^{W^-\to f_3\bar f_4}(\lambda_-;\tau_4,\bar\tau_3) 
     \nonumber \\
  &&~~~ +{\cal H}_{Born}^{e^+e^-\to W^+W^-}(\sigma_+,\sigma_-;\lambda_+,\lambda_-)
      \delta {\cal H}^{W^+\to f_1\bar f_2}(\lambda_+;\tau_1,\bar\tau_2)
      {\cal H}_{Born}^{W^-\to f_3\bar f_4}(\lambda_-;\tau_4,\bar\tau_3) 
      \nonumber \\
  &&~~~ +{\cal H}_{Born}^{e^+e^-\to W^+W^-}(\sigma_+,\sigma_-;\lambda_+,\lambda_-)
      {\cal H}_{Born}^{W^+\to f_1\bar f_2}(\lambda_+;\tau_1,\bar\tau_2)
      \delta{\cal H}^{W^-\to f_3\bar f_4}(\lambda_-;\tau_4,\bar\tau_3),
      \nonumber
\eqa
and 
\bq
   K_\pm \equiv k_\pm^2-M_W^2+iM_W\Gamma_W,  \label{4.3}
\eq
where $\sqrt{k_\pm^2}$ is the invariant mass of the fermion pair.  
All the helicity amplitudes
appearing in (\ref{4.2}) are evaluated on the mass-shell of 
the $W^\pm$-bosons. 
The first line corresponds to the Born amplitude, 
while the second to the fourth line represent
the one-loop corrections to  the production and 
the $W^\pm$-decay subprocesses.
The spin correlation is fully taken into account in 
formulation (4.1).\par
     In order to regularize the infrared singularity, we include 
the soft-photon contribution in  the virtual-loop corrections.\par
    In terms of the helicity amplitudes (4.2),  the differential
cross section is given by
\bqa
   d\sigma& =& {1\over {2s}} (2\pi)^{-7}{1\over{512}} 
           {\sum} ^\prime \int  d\cos\theta dk_+^2dk_-^2
              d\cos\hat\theta_2d\hat\varphi_2d\cos\hat\theta_4d\hat\varphi_4
               \nonumber\\
     && {{\vert \vec{k_\pm}\vert}\over {E_{beam}}}
        {{ \vert {\cal H }(e^+(\sigma_+)e^-(\sigma_-) \to 
                 W^+_{\lambda_+}(f_1(\tau_1)\bar f_2(\bar\tau_2))
                 W^-_{\lambda_-}(f_4(\tau_4)\bar f_3(\bar\tau_3)))
           \vert^2 }\over{ [(k_+^2-M_W^2)^2+\Gamma_W^2 M_W^2]  
               [(k_-^2-M_W^2)^2+\Gamma_W^2 M_W^2] }}, \label{4.4}
\eqa
where ${\sum}^\prime$ indicates the spin sum over the final fermions 
as well as the spin average of the initial $e^+e^-$ pair, and
it is understood that only the corrections of up to the order $g^2$ are 
retained in the squares of the helicity amplitudes.  
The cosine of the W-production angle is denoted  by $\cos\theta$, and 
$\hat\theta_2(\hat\theta_4)$ and 
$\hat\varphi_2(\hat\varphi_4)$ are  the decay angles of fermion 2 (4) 
in the rest frame of the $W^+(W^-)$-boson. 
For further details on the notation and
the kinematics, we refer to Appendix A.
The magnitude of the  three-momentum of the $W^\pm$ in the $e^+e^-$ 
center-of-mass frame is
denoted by $\vert \vec{ k_\pm}\vert$, where 
\bq
    \vert \vec{k_\pm}\vert^2 = {1\over{4s}}(s^2+ k_+^4+ k_-^4 -2s(k_+^2+k_-^2)
                                     -2k_+^2k_-^2). \label{4.5}
\eq
\par
     According to the analysis and discussion of Section 3, 
the range of the integration over the fermion-pair masses  squared, 
$k_\pm^2$ , is to be restricted 
in order to sufficiently reduce non-double-resonance
four-fermion  production.  While too small an integration
range will strongly reduce event rates, too large a range
will increase background.  We find that the five-$\Gamma_W$ choice (3.2)
constitutes a reasonable compromise.\par
The seven-fold non-trival 
integral (\ref{4.4}) can be split into two integrals, 
the two-fold integral over the Breit-Wigner denominators defined by 
\bqa
     I(s, M_W, \Gamma_W)& \equiv& \int_D dk_+^2 dk_-^2 
      {{\vert\vec{ k_\pm}\vert}\over {E_{beam}}} \nonumber \\
   && {1\over{ [(k_+^2-M_W^2)^2+\Gamma_W^2 M_W^2]  
               [(k_-^2-M_W^2)^2+\Gamma_W^2 M_W^2] }}, \label{4.6}
\eqa
with the integration region $D$  restricted  by the cut (3.2), and 
the remaining five-fold integral. The cross section  (\ref{4.4}) 
then becomes
\bqa
   d\sigma& =& {1\over {2s}} (2\pi)^{-7}{{I(s,M_W,\Gamma_W)}\over{512}} 
    {\sum} ^\prime \int d\cos\theta d\cos\hat\theta_2d\hat\varphi_2
                        d\cos\hat\theta_4d\hat\varphi_4 
           \nonumber\\
  && ~~~~~~~~~~\vert 
     {\cal H}_{Born}^{e^+e^-\to W^+W^-}{\cal H}_{Born}^{W^+\to f_1\bar f_2}
      {\cal H}_{Born}^{W^-\to f_3\bar f_4} \nonumber \\
  &&~~~~~~~~~~~
      +\delta{\cal H}^{e^+e^-\to W^+W^-}{\cal H}_{Born}^{W^+\to f_1\bar f_2}
      {\cal H}_{Born}^{W^-\to f_3\bar f_4} \nonumber \\
  &&~~~~~~~~~~~
      +{\cal H}_{Born}^{e^+e^-\to W^+W^-}\delta {\cal H}^{W^+\to f_1\bar f_2}
      {\cal H}_{Born}^{W^-\to f_3\bar f_4} \nonumber \\
  &&~~~~~~~~~~~
      +{\cal H}_{Born}^{e^+e^-\to W^+W^-}{\cal H}_{Born}^{W^+\to f_1\bar f_2}
      \delta{\cal H}^{W^-\to f_3\bar f_4} \vert^2. \label{4.7}
\eqa
The integral $I$ thus  plays the role of a weight factor originating 
from  the Breit-Wigner 
propagators.  Note that the integral $I$ depends on $s$ through 
$ \vert \vec{k_\pm} \vert/E_{beam}$. \par
     The decay width of the $W^\pm$-boson appearing in (\ref{4.7}) 
is computed by {\tt GRACE} \cite{Grace}.  The resulting width
\bq
    \Gamma_W = 2.046 GeV, \label{4.8}
\eq
includes the  full one-loop radiative corrections with soft and 
hard bremsstrahlung of photon and gluon. 
The masses of the particles appearing in the loop
calculation are listed in Appendix A.  The numerical values
of the integral $I$ for the cut (3.2) on $k_\pm^2$  are
given in Table 2.  The width $\Gamma_W$ entering the Born value in
table 2 is given by $\Gamma_W=1.942$ GeV, while the one-loop result 
is based on (\ref{4.8}).\par

\begin{table}
\begin{tabular}{|r|l|l|}\hline
    $E_{beam}(\rm GeV)$  &  Born($GeV^{-4}$) & one-loop($GeV^{-4}$) \\ \hline
    200 & 3.2646$\times 10^{-4}$ & 2.941$\times 10^{-4}$  \\ \hline
    300 & 3.4343$\times 10^{-4}$ & 3.094$\times 10^{-4}$  \\ \hline
    400 & 3.4917$\times 10^{-4}$ & 3.146$\times 10^{-4}$  \\ \hline
    500 & 3.5180$\times 10^{-4}$ & 3.169$\times 10^{-4}$  \\ \hline
    600 & 3.5322$\times 10^{-4}$ & 3.182$\times 10^{-4}$  \\ \hline
    700 & 3.5407$\times 10^{-4}$ & 3.190$\times 10^{-4}$  \\ \hline
    800 & 3.5462$\times 10^{-4}$ & 3.195$\times 10^{-4}$  \\ \hline
    900 & 3.5500$\times 10^{-4}$ & 3.198$\times 10^{-4}$  \\ \hline
   1000 & 3.5527$\times 10^{-4}$ & 3.201$\times 10^{-4}$  \\ \hline
\end{tabular}

\medskip
\noindent{Table 2}. The weight factor $I(s,M_W, \Gamma_W)$ for 
the cut on $k_+^2$ and $k_-^2$  defined by (3.2).\par
\end{table}


   For the one-loop  helicity amplitudes of
$e^+e^- \to W^+W^-$ in (\ref{4.7}),
we will use the high-energy-Born-form approxiamtion (HEBFA) of 
Ref.\cite{Kuschi}.  In the HEBFA, the helicity amplitudes take 
the form \cite{Kuschi}
\bq
    {\cal H}(\sigma, \lambda_+,\lambda_-)
  = S_I^{(\sigma)}{\cal M}_I(\sigma, \lambda_+,\lambda_-)\delta_{\sigma,-} 
 +  S_Q^{(\sigma)}{\cal M}_Q(\sigma, \lambda_+,\lambda_-),\label{4.9}
\eq
where the invariant amplitudes $ S_I^{(-)}(s,t,\Delta E)$ and  
$S_Q^{(\pm)}(s,t,\Delta E)$ contain the one-loop virtual corrections as
well as the soft-photon radiation with soft-photon-energy cut  $\Delta E$.
It is worth noting that the analytical formulae for the one-loop
invariant amplitudes in (\ref{4.9}) are very simple and  
can be written down in a few lines \cite{Kuschi}.
For definiteness and completeness, we will also compare with the result
obtained by using the full one-loop-corrected amplitudes.
Their analytical expressions were calculated in 
Refs.\cite{Boehm,Fleischer}.  For our numerical evaluation we will
use the results of an independent calculation by one of us 
\cite{Kuroda}.  The numerical results of Ref.\cite{Kuroda}
agree well with the ones of Ref.\cite{Boehm}.\par

     Concerning the decay amplitudes for $W^+\to f_1 \bar f_2$
and $W^-\to f_4 \bar f_3$
we can safely neglect fermion masses with a discrepancy of less 
than 0.3\% \cite{Ditt2}. In this approximation, up to  one-loop order, 
the decay amplitudes can be expressed as (see also (6.18) of 
Ref.\cite{Denner2}),
\bqa
     {\cal H}^{W^+\to f_1\bar f_2}(\lambda_+; \tau_1, \bar\tau_2)&=& 
     {g\over{2\sqrt 2}}G^{(+)}( p_{W^+}, p_1, p_2)
     {\cal M}^{(+)}(\lambda_+;\tau_1,\bar\tau_2), \nonumber \\
     {\cal H}^{W^-\to f_4\bar f_3}(\lambda_-; \tau_4, \bar\tau_3)& =& 
     {g\over{2\sqrt 2}}G^{(-)}( p_{W^-}, p_4, p_3)
     {\cal M}^{(-)}(\lambda_-;\tau_4,\bar\tau_3),\label{4.10}
\eqa
where  $\lambda_\pm$, $\tau_1$,  $\bar \tau_2$, $\bar\tau_3$ and $\tau_4$
are the helicities of $W^\pm$, and twice of the helicities of 
$f_1$, $\bar f_2$, $\bar f_3$ and $f_4$, respectively.
The basic amplitudes ${\cal M}^{(\pm)}$ are defined by
\bqa
   {\cal M}^{(+)} (\lambda_+;\tau_1, \bar\tau_2)= 
         \bar u(p_1, -) \gamma_\mu(1-\gamma_5) v(p_2,+)
                            \epsilon_+^\mu(p_{W^+},\lambda_+)
         \delta_{\tau_1, -}\delta_{\bar\tau_2,+}, \nonumber \\
   {\cal M}^{(-)} (\lambda_-;\tau_4, \bar\tau_3)= 
         \bar u(p_4, -) \gamma_\mu(1-\gamma_5) v(p_3,+)
                            \epsilon_-^\mu(p_{W^-},\lambda_-)
         \delta_{\tau_4, -}\delta_{\bar\tau_3,+},\label{4.11}
\eqa
in the massless limit, while the  invariant amplitude $G^{(\pm)}$ 
contains all the dynamic information of the process.  
The invariant amplitude   $G^{(\pm)}$ is normalized to unity at Born level
\bq
       G^{(\pm)}_{Born} =1.    \label{4.12}
\eq\par
     At one-loop level, $G^{(\pm)}$ receives contributions from 
virtual diagrams,
the counterterm lagrangian and from the soft-photon radiation.
Separating the dominant fermion contribution,
it is expressed as,
\bq
   G^{(\pm)} = 1 + \Delta \alpha(M_W^2)- {{c_W^2}\over{2s_W^2}}\Delta\rho
         + {1\over 2}\Delta_{LL} + G^{(\pm, rest)}, \label{4.13}
\eq
where the first term is the Born contribution, 
\bqa
   \Delta\alpha(s)&=& {{\alpha}\over{3\pi}}\sum_f Q_f^2\log{s\over{m_f^2}}, \\
   \Delta\rho     &=& {{g^2}\over{16\pi^2}} {{3m_t^2}\over{4M_W^2}}, \\
   \Delta_{LL}&=& {{2e^2}\over{16\pi^2}}
      [\{ Q_1^2\log{{M_W^2}\over{m_1^2}}+Q_2^2\log{{M_W^2}\over{m_2^2}}
         -Q_1^2-Q_2^2-1 \}\log{{(2\Delta E)^2}\over{M_W^2}} \nonumber\\
        & &~~~~~~~+({3\over 2} -\log{{4p_{10}^2}\over{M_W^2}})
                     Q_1^2\log{{M_W^2}\over{m_1^2}}
                 +({3\over 2} -\log{{4p_{20}^2}\over{M_W^2}})
                     Q_1^2\log{{M_W^2}\over{m_2^2}}  ],\label{4.16}
\eqa
and $G^{(\pm,rest)}$ is the remaining part.  
The expression for $\Delta_{LL}$, due to the soft-photon radiation,  
should be compared with (3.7) of Ref.\cite{Kuschi}
for $e^+e^- \to W^+W^-$.  In the present case, we have to evaluate the
invariant amplitudes $G^{(\pm)}$  not in the rest frame of the 
$W^\pm$-boson,  but in the laboratory frame ( the c.m. frame of the 
$e^+e^-$-pair), and, consequently, for $\Delta E$ we have to use 
the same numerical value as
the one used in the $W$-pair production subprocess.\par
     The full expressions for the basic amplitude, ${\cal M}^{(\pm)}$,
and the remaining part, $G^{(\pm, rest)}$, are given in Appendices 
A and B, respectively. 
     
\setcounter{equation}{0}
\section{Hard-photon radiation.}
     In DPA, the hard photon radiation can be treated in a way that is
analogous to the above treatment of $e^+e^-\to W^+W^-\to 4$ fermions.
Since interference is negligible, the cross section in DPA can be described
by \cite{W.Been},\footnote{Hard-photon radiation in $e^+e^-\to
4$~fermions$+\gamma$ at tree level, not based on a DPA was recently
treated in Refs.\cite{Ditt1,Jeger}}
\bqa
     d\sigma &=& {1\over {2s}}
            \bigg |{{ {\cal H}^{e^+e^-\to W^+W^-\gamma}
                      {\cal H}^{W^+ \to f_1\bar f_2}
                      {\cal H}^{W^- \to f_4\bar f_3}}\over{K_+K_-}}
            \bigg|^2  {{dk_+^2}\over {2\pi}}{{dk_-^2}\over {2\pi}}
            d\Gamma_{\rm pro}^\gamma d\Gamma_{\rm dec}^+
            d\Gamma_{\rm dec}^- \nonumber \\
            &+ & {1\over {2s}}
            \bigg |{{{\cal H}^{e^+e^-\to W^+W^-}
                     {\cal H}^{W^+ \to f_1\bar f_2\gamma}
                     {\cal H}^{W^- \to f_4\bar f_3}}\over{K_{+\gamma} K_-}}
            \bigg|^2 {{dk_+^2}\over {2\pi}}{{dk_-^2}\over {2\pi}}
            d\Gamma_{\rm pro} d\Gamma_{\rm dec}^{+\gamma}
            d\Gamma_{\rm dec}^- \nonumber \\
            &+ & {1\over {2s}}
            \bigg |{{{\cal H}^{e^+e^-\to W^+W^-}
                     {\cal H}^{W^+ \to f_1\bar f_2}
                {\cal H}^{W^- \to f_4\bar f_3\gamma}}\over{K_+K_{-\gamma}}}
            \bigg|^2 {{dk_+^2}\over {2\pi}}{{dk_-^2}\over{2\pi}}
            d\Gamma_{\rm pro} d\Gamma_{\rm dec}^+
            d\Gamma_{\rm dec}^{-\gamma} \label{5.1}
\eqa
where
\bqa
     d\Gamma_{\rm pro}&=&{1\over{(2\pi)^2}}\delta^4(k_1+k_2-k_+-k_-) 
                         {{d^3k_+}\over{2k_{+0}}}{{d^3k_-}\over{2k_{-0}}},
     \nonumber \\   
     d\Gamma_{\rm dec}^+&=&{1\over{(2\pi)^2}}\delta^4(k_+-p_1-p_2) 
                         {{d^3p_1}\over{2p_{10}}}{{d^3p_2}\over{2p_{20}}},
     \nonumber \\   
     d\Gamma_{\rm dec}^-&=&{1\over{(2\pi)^2}}\delta^4(k_--p_3-p_4) 
                         {{d^3p_3}\over{2p_{30}}}{{d^3p_4}\over{2p_{40}}},
     \nonumber \\   
     d\Gamma_{\rm pro}^\gamma&=&{1\over{(2\pi)^5}}
                         \delta^4(k_1+k_2-k_+-k_--k_\gamma) 
                         {{d^3k_+}\over{2k_{+0}}}{{d^3k_-}\over{2k_{-0}}}
                         {{d^3k_\gamma}\over{2k_{\gamma 0}}},
     \nonumber \\   
     d\Gamma_{\rm dec}^{+\gamma}&=&{1\over{(2\pi)^5}}
                         \delta^4(k_+-p_1-p_2-k_\gamma) 
                         {{d^3p_1}\over{2p_{10}}}{{d^3p_2}\over{2p_{20}}}
                         {{d^3k_\gamma}\over{2k_{\gamma 0}}},
     \nonumber \\   
     d\Gamma_{\rm dec}^{-\gamma}&=&{1\over{(2\pi)^5}}
                         \delta^4(k_--p_3-p_4-k_\gamma) 
                         {{d^3p_3}\over{2p_{30}}}{{d^3p_4}\over{2p_{40}}}
                         {{d^3k_\gamma}\over{2k_{\gamma 0}}},
     \label{5.3}  
\eqa
and
\bqa
     K_+ &=& (p_1+p_2)^2-M_W^2+iM_W\Gamma_W, \nonumber \\
     K_- &=& (p_3+p_4)^2-M_W^2+iM_W\Gamma_W, \nonumber \\
     K_{+\gamma} &=& (p_1+p_2+k_\gamma)^2-M_W^2+iM_W\Gamma_W, \label{5.4} \\
     K_{-\gamma} &=& (p_3+p_4+k_\gamma)^2-M_W^2+iM_W\Gamma_W. \nonumber \\
\eqa
In (5.1) the sum over the helicities of the intermediate $W^\pm$-bosons
is implicitly assumed. Therefore,  
the full spin-correlation is incorporated also in the hard-photon
bremsstrahlung process.

\setcounter{equation}{0}
\section{Numerical Results}

     In this section, we present the  results for 
\bq
      e^+e^-\to W^+(u\bar d) W^-( \bar c s) [+\gamma], \label{6.1}
\eq
at one-loop level.  As mentioned, we~ employ one-loop on-shell $W^\pm$-
production~ and -decay amplitudes together with (fixed width)
Breit-Wigner denominators and the cut $M_W-5\Gamma_W \le \sqrt{k^2_\pm}
\le M_W+5\Gamma_W$ on the fermion-pair masses.  The cut strongly
enhances the $W^+W^-$ signal relative to contributions from the 
non-doubly-resonant background that is not taken into account in the
calculation.  From the point of view of four-fermion production
directly observed experimentally, our results for (\ref{6.1}) 
correspond to a double-pole approximation (DPA).\par

     We summarize the main steps of the calculation:\par
i)  For the amplitudes of $e^+e^-\to W^+W^-$ at one-loop level
we use the high-energy Born-form approximation (HEBFA) previously
constructed \cite{Kuschi}.  Even though the comparison
of the HEBFA  with the full one-loop result for $e^+e^- \to W^+W^-$
showed excellent agreement \cite{Kuschi}, we will nevertheless 
also present results  for the reaction (\ref{6.1}) that are based on the
full one-loop amplitudes.  The comparison of the results
based on the HEBFA with the results from the full one-loop amplitudes 
will allow us to quantitatively state the accuracy of the 
HEBFA even when the decay of the W-boson
is included.\footnote{In Ref.\cite{Kuschi}, we 
ignored the W decay.  As different helicities
enter the cross section for 
$e^+e^-\to W^+W^-\to 4$ fermions with different weights, 
it is not a priori guaranteed that the
accuracies excluding and including W decay are identical.}
\par
ii)  For the decay amplitudes, $W^\pm \to f \bar f$, we use the full
one-loop expression summarized in Appendix B.  The fermion mass,
with the exception of mass-singular terms, is
neglected in the calculation of virtual and soft-photon corrections.
\par
iii)  The amplitudes for hard-photon emission are generated by
using the algebraic manipulation program {\tt GRACE} \cite{Grace},
and the necessary numerical integrations are carried out by 
employing the Monte Carlo routine {\tt BASES} \cite{BASES}.
We require the error due to the Monte Carlo integration to be 
less than about 0.1 \%.  The independence of the result under the
variation of the soft-photon cut $\Delta E$ is verified by varing
$\Delta E$ between 1 GeV and 10 GeV.\par

     The calculations are based  on the following input parameters [in
GeV],
\bqa
     M_Z =& 91.187, ~~~~M_W = 80.22, ~~~~~M_H = 200, \cr
     m_u =& 0.062, ~~~~~m_c = 1.5, ~~~~~~~~~~m_t = 175, \cr
     m_d =& 0.083, ~~~~~m_s = 0.215, ~~~~~~m_b = 4.5.
\label{6.2}
\eqa
\par
     First of all, in Table 3 we consider the $\cos\theta$ angular 
distribution of the produced $W$ pairs at a fixed energy that is chosen
as $\sqrt s = 2E_{\rm beam} = 1 $TeV.  For easy reference,
column 1 in Table 3 shows the Born-approximation results for 
$e^+e^-\to W^+W^-$.  The second column of Table 3 
includes the decay of the
W-bosons, obtained by integrating  the Breit-Wigner decay distributions
 over the
restricted region (3.2).  This restriction of the invariant mass
of the fermion pairs leads to a slightly smaller cross section in column
2 than calculated by multiplication of the cross section for 
$e^+e^-\to W^+W^-$ from 
column 1 by the square of the branching ratio, $BR(W\to q \bar q) =
1/3$.\par
\medskip

\small{
\begin{table}
\begin{tabular}{|r|c|c|c|c|r|}\hline
              & $e^+e^-\to W^+W^-$ & $e^+e^-\to W^+(u\bar d)W^-(\bar c s)$
              & \multicolumn{3}{c|}
                {$e^+e^-\to W^+(u\bar d)W^-(\bar c s)+\gamma$} \\ \cline{2-6}
              & Born &  Born & \multicolumn{3} {c|}{one-loop}  \\ \cline{4-6}
  $\cos\theta$   &    &   & HEBFA  &  exact & $\Delta (\%)$ \\ \hline
  0.95& $5.981\times 10^0  $  & $5.827\times 10^{-1}$ &
        $2.900\times 10^{-1}$ & $2.878\times 10^{-1}$ & 0.76 \\ 
  0.9 & $2.785\times 10^0  $  & $2.713\times 10^{-1}$ &
        $1.211\times 10^{-1}$ & $1.208\times 10^{-1}$ & 0.24 \\ 
  0.8 & $1.207\times 10^0  $  & $1,176\times 10^{-1}$ &
        $4.557\times 10^{-2}$ & $4.557\times 10^{-2}$ & 0.00\\
  0.7 & $7.003\times 10^{-1}$ & $6,826\times 10^{-2}$ &
        $2.383\times 10^{-2}$ & $2.385\times 10^{-2}$ & -0.05\\
  0.6 & $4.597\times 10^{-1}$ & $4.483\times 10^{-2}$ &
        $1.437\times 10^{-2}$ & $1.438\times 10^{-2}$ & -0.02\\
  0.5 & $3.246\times 10^{-1}$ & $3.165\times 10^{-2}$ &
        $9.429\times 10^{-3}$ & $9.435\times 10^{-2}$ & -0.06 \\
  0.4 & $2.414\times 10^{-1}$ & $2.352\times 10^{-2}$ &
        $6.570\times 10^{-3}$ & $6.576\times 10^{-3}$ & -0.10 \\
  0.3 & $1.869\times 10^{-1}$ & $1.821\times 10^{-2}$ &
        $4.798\times 10^{-3}$ & $4.808\times 10^{-3}$ & -0.20 \\
  0.2 & $1.497\times 10^{-1}$ & $1.458\times 10^{-2}$ &
        $3.645\times 10^{-3}$ & $3.651\times 10^{-3}$ & -0.17\\
  0.1 & $1.234\times 10^{-1}$ & $1.201\times 10^{-2}$ &
        $2.855\times 10^{-3}$ & $2.861\times 10^{-3}$ & -0.22 \\
  0.0 & $1.041\times 10^{-1}$ & $1.013\times 10^{-2}$ &
        $2.292\times 10^{-3}$ & $2.297\times 10^{-3}$ & -0.23 \\
 -0.1 & $8.941\times 10^{-2}$ & $8.695\times 10^{-3}$ &
        $1.872\times 10^{-3}$ & $1.876\times 10^{-3}$ & -0.22 \\
 -0.2 & $7.766\times 10^{-2}$ & $7.551\times 10^{-3}$ &
        $1.542\times 10^{-3}$ & $1.544\times 10^{-3}$ & -0.16 \\
 -0.3 & $6.773\times 10^{-2}$ & $6.586\times 10^{-3}$ &
        $1.268\times 10^{-3}$ & $1.269\times 10^{-3}$ & -0.05 \\
 -0.4 & $5.883\times 10^{-2}$ & $5.721\times 10^{-3}$ &
        $1.031\times 10^{-3}$ & $1.030\times 10^{-3}$ &  0.06 \\
 -0.5 & $5.036\times 10^{-2}$ & $4.897\times 10^{-3}$ &
        $8.174\times 10^{-4}$ & $8.148\times 10^{-4}$ &  0.31 \\
 -0.6 & $4.188\times 10^{-2}$ & $4.073\times 10^{-3}$ &
        $6.202\times 10^{-4}$ & $6.155\times 10^{-4}$ &  0.77 \\
 -0.7 & $3.305\times 10^{-2}$ & $3.214\times 10^{-3}$ &
        $4.364\times 10^{-4}$ & $4.291\times 10^{-4}$ &  1.70 \\
 -0.8 & $2.360\times 10^{-2}$ & $2.295\times 10^{-3}$ &
        $2.680\times 10^{-4}$ & $2.573\times 10^{-4}$ &  4.14 \\
 -0.9 & $1.333\times 10^{-2}$ & $1.296\times 10^{-3}$ &
        $1.215\times 10^{-4}$ & $1.074\times 10^{-4}$ & 13.19 \\ \hline
\end{tabular}
\par
\medskip
\footnotesize{
\noindent{Table 3.} The angular distribution  of W-pair production 
at the  energy $\sqrt s =2E_{beam} = 1$ TeV in units of $pb$.  
The first column shows  the Born cross section for $e^+e^-\to W^+W^-$. 
The second column shows the results of treating W production and decay
in Born approximation and integrating the Breit-Wigner distribution
over the restricted interval (3.2).
The third and the fourth column are obtained by using the one-loop
amplitudes for production and decay, again, integrating the Breit-Wigner
distribution over the restricted interval (3.2).
A soft-photon cut $\Delta E/E= 0.01$ is used for the one-loop
results.  The HEBFA is used for the third column and the full one-loop
amplitudes are  used for the fourth column.  The last column gives the
results for the relative deviation, $\Delta$, from (\ref{6.3}).
}
\end{table}
}
\normalsize{
     The main result on the angular distribution is contained in the
two one-loop columns of Table 3.  The results are obtained for 
$E_{\rm beam}=500$GeV and an
infrared cut-off $\Delta E/E=0.01$, i.e. for $\Delta E = 5$ GeV.
As mentioned, the one-loop amplitudes for W-pair production 
are supplemented by one-loop
decay amplitudes, and the Breit-Wigner distributions  are integrated over
the restricted invariant-mass interval of 5 times the W-boson width,
$\Gamma_W$, according to (3.2).
The percentage deviations
\bq
     \Delta = {{  {{d\sigma}\over{d\cos\theta}}({\rm HEBFA}) -
                  {{d\sigma}\over{d\cos\theta}}({\rm exact}) } \over
               {  {{d\sigma}\over{d\cos\theta}}({\rm exact}) }}, 
\label{6.3}
\eq
according to the last column of Table 3, stay below 3 per mill in 
most of the angular interval.  
As expected from the behavior of the three
invariant amplitudes entering the HEBFA \cite{Kuschi}, 
the deviations between the approximation
and the full one-loop results become largest (order  of several 
percent) for very forward and backward angles. 
Nevertheless, the calculationally very fast and intuitively simple 
HEBFA, for most of the range of the production angle, yields 
an excellent approximation of the full
one-loop results for the differential cross section.\par

     We turn to the total cross section as a function of the 
$e^+e^-$ energy, obtained by integration over the angular 
distribution of the W-boson pair.  Tables 4 and 5, respectively,
show the cross sections obtained by integration over the full
angular range of $0^\circ<\theta<180^\circ$ and over the restricted
range of $ 10^\circ<\theta <170^\circ$.   As the forward peaking
of the cross section increases strongly with energy, the angular cut
with increasing energy leads to an increasingly  stronger reduction of the 
integrated cross section.  This is seen when comparing the cross
sections in Tables 4 and 5.  As a result of this angular cut, the 
accuracy of the HEBFA is strongly increased.  The difference,
\bq
     \Delta = {{\sigma({\rm HEBFA}) - \sigma({\rm exact})}
                                \over{\sigma({\rm exact}) }}, 
\label{6.4}
\eq
also shown in Tables 4 and 5 , by removing the very forward- and
backward- production angles is diminished by an order of magnitude,
from about $3\%$ to about $0.3\%$.  The increase of accuracy is
expected.
As noted in connection with the results in Table 3, 
in the very forward and backward directions, the
invariant amplitudes entering the HEBFA are less accurate \cite{Kuschi}.
\par
}

\small{
\begin{table}
\begin{tabular}{|r|c|c|c|c|c|}\hline
              & $e^+e^-\to W^+W^-$ & $e^+e^-\to W^+(u\bar d)W^-(\bar c s)$
              & \multicolumn{3}{c|}
                {$e^+e^-\to W^+(u\bar d)W^-(\bar c s)+\gamma$} \\ \cline{2-6}
               & Born    &  Born  &\multicolumn{3}{c|}{one-loop}\\ \cline{4-6}
 $E_{\rm beam}$&         &        & HEBFA  &  exact  & $\Delta(\%)$ 
                                                     \\ \hline
  200 & 8.698  & 0.8467  & 0.8718 & 0.8794 &-0.9  \\ 
  300 & 5.091  & 0.4958  & 0.5275 & 0.5262 & 0.2  \\ 
  400 & 3.384  & 0.3295  & 0.3575 & 0.3533 & 1.2  \\ 
  500 & 2.433  & 0.2370  & 0.2602 & 0.2556 & 1.8  \\ 
  600 & 1.844  & 0.1795  & 0.1996 & 0.1951 & 2.3  \\ 
  700 & 1.452  & 0.1413  & 0.1584 & 0.1543 & 2.7  \\ 
  800 & 1.177  & 0.1145  & 0.1292 & 0.1259 & 2.6  \\ 
  900 & 0.9750 & 0.09485 & 0.1080 & 0.1044 & 3.4  \\ 
 1000 & 0.8228 & 0.08010 & 0.0915 & 0.0881 & 3.9  \\ \hline
\end{tabular}
\par
\medskip
\footnotesize{
\noindent{ Table 4.} The energy dependence of the $(u\bar d)(\bar c s)$- 
production cross section in DPA. The second column is the Born cross section,
while the third column gives  the  one-loop cross section 
including hard-photon radiation. The deviation, $\Delta$, according
to (\ref{6.4}), quantifies the discrepancy between the HEBFA and the
full one-loop results.  \par
}
\end{table}

\vspace{0.5 cm}
\begin{table}
\begin{tabular}{|r|c|c|c|c|c|}\hline
              & $e^+e^-\to W^+W^-$ & $e^+e^-\to W^+(u\bar d)W^-(\bar c s)$
              & \multicolumn{3}{c|}
                {$e^+e^-\to W^+(u\bar d)W^-(\bar c s)+\gamma$} \\ \cline{2-6}
               & Born    &  Born   &\multicolumn{3}{c|}{one-loop}\\ \cline{4-6}
 $E_{\rm beam}$&         &         & HEBFA   &  exact  & $\Delta(\%)$
                                                        \\ \hline
  200 & 6.724  & 0.6561  & 0.6746  & 0.6794  &-0.71  \\
  300 & 3.042  & 0.2964  & 0.3109  & 0.3109  & 0.00  \\
  400 & 1.695  & 0.1654  & 0.1725  & 0.1721  & 0.23  \\
  500 & 1.077  & 0.1051  & 0.1085  & 0.1082  & 0.28  \\
  600 & 0.7440 & 0.07262 & 0.07405 & 0.07383 & 0.30  \\
  700 & 0.5449 & 0.05318 & 0.05349 & 0.05334 & 0.28  \\
  800 & 0.4162 & 0.04063 & 0.04027 & 0.04015 & 0.30  \\
  900 & 0.3284 & 0.03205 & 0.03136 & 0.03127 & 0.29  \\
 1000 & 0.2657 & 0.02593 & 0.02505 & 0.02498 & 0.28  \\ \hline
\end{tabular}
\par
\medskip
\footnotesize{
\noindent{ Table 5.} As Table 4, but with 
a restriction on the $W^+W^-$ production angle that is given by 
$10^\circ< \theta <170^\circ$. 
}
\end{table}
}

\footnotesize{
\begin{table}
\begin{tabular}{|c|c|c|c|c|c|c|c|c|c|}\hline
 $E_{\rm beam}$ & 200 & 300 & 400 & 500 & 600 & 700 & 800 & 900 & 1000
\\ \hline
 $\sigma(E_\gamma<10)$ &  0.4787 & 0.1928  & 0.0958  & 0.0544 & 0.03361
                       & 0.02195 & 0.01486 & 0.01033 & 0.00733 
 \\
 $\sigma(E_\gamma>10)$ &  0.2007 & 0.1181  & 0.0763  & 0.0538 & 0.04022
                       & 0.03139 & 0.02529 & 0.02094 & 0.01765
\\ \hline
 $\sigma$              &  0.6794& 0.3109  & 0.1721  & 0.1082  & 0.07383
                       & 0.05334 & 0.04015 & 0.03127 & 0.02498 
\\ \hline
\end{tabular}
\par
\medskip
\noindent{Table 6.} The total cross  section for $e^+e^- \to
W^+(u \bar d)W^-(\bar c s)\gamma$ is split into two parts
according to the photon energy.  The second line shows the total
cross section with photon energy $E_\gamma<10$ GeV, while the third
line shows the cross section with $E_\gamma>10$ GeV.  The fourth 
line is the sum of these two, namely the total cross section. 
As for the results in Table 5, a cut in the $W^+W^-$ production angle of
$10^\circ < \theta < 170^\circ$ is imposed.
\end{table}
}
\normalsize{
     In the high-energy limit, applying
the cut $10^\circ<\theta<170^\circ$, the total cross section
deviates from the Born result only by a few percent.  On the other
hand, the cross section is strongly dominated by hard-photon
radiation.  This is shown in table 6.  At a beam energy of
$E_{\rm beam}=200$ GeV, a fraction of $30\%$ of the cross section
contains
a hard photon of energy $E_\gamma>10$ GeV, and this fraction rises
strongly with increasing beam energy. At an energy of $E_{\rm beam}= 1$
TeV,
a photon-energy cut of $E_\gamma >10$ GeV removes almost $75\%$
of the cross section.\par

     We finally add a very brief remark on the accuracy to be aimed at
in future experiments in order to obtain a meaningful test of the
non-Abelian structure of the electroweak  theory.
The non-Abelian structure enters both at tree level as well as at one
loop.  In Ref.\cite{Kuschi}, the invariant amplitudes of the HEBFA in
(4.9)
were represented as a sum of two parts.  The first part\footnote{
Compare (3.3) and (3.4) as well as Figs.1 to 3 in Ref.\cite{Kuschi}.}
contains only the leading fermion-loop corrections due to the
light leptons and quarks and the heavy top quark, as well as the
initial state radiation in leading-log approximation.
The second additive part needs the full machinary of the
non-Abelian electroweak theory.
Taking into account only the aforementioned fermion loops,
one finds that the cross section becomes larger 
than the full one-loop - corrected one. Quantitatively, at 
$\sqrt s = 1$ Tev, with the fermion loops alone, the total 
cross- section
becomes 0.119$pb$, a value almost 11\% larger than
the corresponding result of 0.108$pb$ from table 5, while at $\sqrt s 
=2$ Tev , one finds 0.030 $pb$, a value about 20\% larger than the
corresponding one of 0.025 $pb$.  An analogy to QED will be helpful 
to understand the negative sign of additional (primarily)
bosonic loop corrections. In QED, photon exchange leads to an infrared
divergence that is cancelled by (necessarily) positive soft-photon
radiation. A similar cancellation mechanism involving (necessarily) 
positive $Z_0$ emission, at ultra-high energies, will 
become relevant in the present case,
thus allowing to understand the negative sign of the bosonic loop 
corrections. Their fairly large magnitude is the result of the
log-squared terms appearing in these corrections 
\cite{Been,Kuschi}. 
Accuracies of the 
order of magnitude of 10\% to 20\% will accordingly allow one to
"see"  the non -Abelian loops.

\vspace{ 1 cm}

\section{Conclusions}
\normalsize{
    The production of four fermions in $e^+e^-$ annihilation at high
energies is dominated by the production and subsequent decay of
two W-bosons.  Our estimate at tree level shows that background
contributions, wherein one of the fermion pairs, or both of them, do not
originate from W decay, are of the order of $5\%$ of the
cross section in the high-energy region ($400 GeV\lsim \sqrt s \lsim 2 TeV$)
under consideration.
Restricting the masses of the fermion pairs to the vicinity of the
W-boson mass, however, removes the background apart from  
a negligible amount of less than 0.3\% for the $(u \bar d)(\bar c s)$
channel and less than 0.9\% for the $(u \bar d)(e\bar\nu_e)$ channel,
the precise value depending on the energy being chosen.
It is accordingly sufficient to concentrate on $e^+e^-
\to W^+W^-\to 4$ fermions and ignore background contributions 
in a refined calculation at one-loop level, even more so, as 
in four-fermion production the main interest lies in the 
test of the  non- Abelian gauge-boson interaction of the 
electroweak theory.
With respect to four-fermion production, 
evaluating $e^+e^-\to W^+W^-\to 4$ fermions  with appropriate cuts on
the invariant masses of the fermion pairs amounts to a double-pole
approximation (DPA).\par

     We have presented results for $e^+e^-\to W^+W^-\to (u \bar d)
(\bar c s)+\gamma$ at one-loop order in the
high-energy limit.  The results explicitly demonstrate that the HEBFA
yields excellent results for the W-pair angular
distribution, and for the total production cross section as well,
if the very forward and backward regions are excluded.  
The results are very satisfactory for theoretical as well as practical
reasons.  Theoretically, the HEBFA is conceptually simple, as it
includes all
relevant virtual one-loop corrections within  three invariant amplitudes
that replace the weak and the electromagnetic coupling appearing 
at tree level.  The HEBFA is of much practical importance, as the
necessary computer time is strongly reduced 
with respect to a calculation that employs the full-one-loop
results.\par

     We finally gave a rough estimate of the accuracy future experiments
are to aim at, in order to test the non-Abelian
(loop) structure of the electroweak theory.

\section{Acknowledgement}
   We would like to thank S. Dittmaier and  J. Fujimoto for 
useful discussions.
}

\newpage
\setcounter{section}{1}
\setcounter{equation}{0}
\renewcommand{\theequation}{\Alph{section}.\arabic{equation}}
\section*{Appendix A. Notations and the basic amplitude for $W^\pm$ decay}
   In this Appendix, we show the explicit expression for the basic decay
amplitudes ${\cal M}^{(\pm)}$ of the $W^\pm$-boson defined by (\ref{4.11}). 
Since we are in the laboratory  frame (i.e. the $e^+e^-$ center-of-mass
frame) in which  the produced $W^\pm$ 
has a large energy, the decay amplitude explicitly depends on the four 
momentum of the $W^\pm$-boson through the soft photon energy cut.
     In accordance with the momentum assignment used in Refs.\cite{Kuschi}
and \cite{FJZ}, 
we define the production angles and the corresponding 
$W^\pm$ polarization vectors\footnote 
{This convention of polarization vectors
is different from the one used by Ref.\cite{Been}, in which $\epsilon_-(-)$, 
$\epsilon_+(0)$ and $\epsilon_+(-)$ have opposite sign.} as follows, 
\bqa
     e^-:~k_1^\mu&=& E(1, 0,0,1), ~~~~~~~~~~~~~~~~~~~~~~~~
     e^+:~k_2^\mu =  E(1, 0,0,-1),   \label{A.1} \\
     W^-:~k_-^\mu&=& E_-(1, \beta_-\sin\theta, 0,\beta_-\cos\theta), ~~~~~
     W^+:~k_+^\mu= E_+(1, -\beta_+\sin\theta, 0,-\beta_+\cos\theta),
          \nonumber
\eqa
with
\bq
          E_\pm  = {{ s + k_\pm^2 -k_\mp^2}\over {2\sqrt s } },~~~~~~
      \beta_\pm  = \sqrt{1-{{k_\pm^2}\over{E_\pm^2}} }. \label{A.2}
\eq
\par
   The decay angles are defined in the laboratory frame as
\bqa
     p_i^\mu &=& E_i(1,\sin\theta_i\cos\phi_i,\sin\theta_i\sin\phi_i,
                    -\cos\theta_i),~~~~~~~~~ (i=1,2) \nonumber\\
     p_j^\mu &=& E_j(1,\sin\theta_j\cos\phi_j,\sin\theta_j\sin\phi_j,
                    \cos\theta_j),~~~~~~~~~~ (j=3,4)  \label{A.3}
\eqa
and in the rest frame of the $W^\pm$-boson as
\bqa
     p_i^\mu &=& {{\sqrt{k_+^2}}\over 2} 
                 (1,\sin\hat\theta_i\cos\hat\varphi_i,
                    \sin\hat\theta_i\sin\hat\varphi_i,
                    -\cos\hat\theta_i),~~~~~~~~~ (i=1,2) \nonumber\\
     p_j^\mu &=& {{\sqrt{k_-^2}}\over 2} 
                 (1,\sin\hat\theta_j\cos\hat\varphi_j,
                    \sin\hat\theta_j\sin\hat\varphi_j,
                    \cos\hat\theta_j).~~~~~~~~~~ (j=3,4)  \label{A.4}
\eqa
In the latter frame, the phase space boundary becomes trivial,
\bq
     -1\le \cos\hat\theta_i \le +1, ~~~~~0\le\hat\varphi_i \le 2\pi. \label{A.5}
\eq     
     After a simple algebra, we find that the energies of the decay products 
in the laboratory frame are related to those of the W-rest frame by     
\bq
    E_i = {{E_\pm}\over 2} (1+\beta_\pm\cos\hat\theta_i),~~~~~~~(i=1,4)
\label{A.6}
\eq
and  the decay angles are related as follows
\bqa
  \left\lgroup\matrix{ \sin\theta_i\cos\varphi_i \cr
                       \sin\theta_i\sin\varphi_i \cr
                       \mp \cos\theta_i }\right\rgroup   =
  {1\over{\beta_\pm+\cos\hat\theta_i}}  
  \left\lgroup\matrix{ \cos\theta & 0 & \sin\theta \cr
                         0        & 1 &  0 \cr
                     - \sin\theta & 0 & \cos\theta }\right\rgroup
  \left\lgroup\matrix{ \sin\hat\theta_i\cos\hat\varphi_i \cr
                       \sin\hat\theta_i\sin\hat\varphi_i \cr
                       \mp(\beta_\pm + \cos\hat\theta_i)}
                       \right\rgroup.
\label{A.7}
\eqa

     The polarization vector of the on-shell $W^\pm$ with  momenta
specified by (\ref{A.1}) are given by
\bqa
  \epsilon_-^\mu(0)&=& {{2M_W}\over{\sqrt s}}(\beta, \sin\theta,0,\cos\theta),
    ~~~~~~~~~~
  \epsilon_+^\mu(0)= {{2M_W}\over{\sqrt s}}(-\beta, \sin\theta,0,\cos\theta),
      \label{A.8}\\
  \epsilon_-^\mu(\pm)&=&\mp {1\over{\sqrt 2}}(0,\cos\theta,\pm i,-\sin\theta),
     ~~~~
  \epsilon_+^\mu(\pm)=\mp {1\over{\sqrt 2}}(0,- \cos\theta,\pm i,\sin\theta),
  \nonumber
\eqa
with
\bq
      \beta  = \sqrt{1-{{4M_W^2}\over s} }. \label{A.9}
\eq
     The non-vanishing basic amplitudes for the on-shell $W^\pm$ decay 
are given by
\bqa
    {\cal M}^{(-)} (\lambda; -,+) &=&\sqrt{4E_3E_4}\bar\varphi_-(p_4)
        [\sigma_\mu\epsilon_-^\mu(\lambda)]\chi_+(p_3), \nonumber \\ 
    {\cal M}^{(+)} (\bar\lambda; -,+)&= &\sqrt{4E_1E_2}\bar\varphi_-(p_1)
        [\sigma_\mu\epsilon_+^\mu(\bar\lambda)]\chi_+(p_2), \label{A.10} 
\eqa
where
\bqa
     \sigma_\mu\epsilon_-^\mu(\lambda) &=& \cases{
        \mp{1\over{\sqrt 2}}\left\lgroup \matrix{
            -\sin\theta & \cos\theta\pm 1 \cr
            \cos\theta\mp 1 & \sin\theta } \right\rgroup, & $\lambda=\pm$\cr
        \gamma\left\lgroup\matrix {
            \beta+\cos\theta & \sin\theta \cr
            \sin\theta & \beta-\cos\theta } \right\rgroup, & $\lambda=0$\cr}
      \nonumber\\
     \sigma_\mu\epsilon_+^\mu(\bar\lambda) &=& \cases{
        \mp{1\over{\sqrt 2}}\left\lgroup \matrix{
            \sin\theta & -\cos\theta\pm 1 \cr
            -\cos\theta\mp 1 & -\sin\theta } \right\rgroup, 
            & $\bar\lambda=\pm$\cr
        \gamma\left\lgroup\matrix {
            -\beta+\cos\theta & \sin\theta \cr
            \sin\theta & -\beta-\cos\theta } \right\rgroup, 
            & $\bar\lambda=0$\cr}.  \label{A.11}
\eqa
and $\varphi_-(p_1)$ and $\chi_+(p_2)$ etc are the two component Weyl
spinors
\bqa
     \varphi_-(p_1)&=& {1\over{\sqrt{2(1-\cos\theta_1)} }}
                       \llgm{-\sin\theta_1 e^{-i\varphi_1}\cr
                             1-\cos\theta_1}\rrgm,  \nonumber\\       
     \chi_+(p_2)&=& {1\over{\sqrt{2(1-\cos\theta_2)} }}
                       \llgm{-\sin\theta_2 e^{-i\varphi_2}\cr
                             1-\cos\theta_2}\rrgm,  \nonumber\\       
     \varphi_-(p_4)&=& {1\over{\sqrt{2(1+\cos\theta_4)} }}
                       \llgm{-\sin\theta_4 e^{-i\varphi_4}\cr
                             1+\cos\theta_4}\rrgm, \nonumber\\       
     \chi_+(p_3)&=& {1\over{\sqrt{2(1+\cos\theta_3)} }}
                       \llgm{-\sin\theta_3 e^{-i\varphi_3}\cr
                             1+\cos\theta_3}\rrgm. \label{A.12}
\eqa
\vskip 0.3 truecm
   For the calculation of radiative corrections, the following values 
are used for our input parameters.
\bqa
     \alpha & =& 1/137.036, \nonumber\\
     M_Z &=& 91.187, ~~~~M_W = 80.22, ~~~~~~M_H = 200, \nonumber \\ 
     m_u &=& 0.062, ~~~~~m_c = 1.5, ~~~~~~~m_t = 175, \nonumber\\
     m_d &=& 0.083, ~~~~~m_s = 0.215, ~~~~~m_b = 4.5, \label{A.13}
\eqa
although, lepton masses and light quark masses are neglected except
for the mass singular terms.
The Born decay width of $W^\pm$ becomes 1.942 GeV, 
while the decay width 
including order $g^2$ and $g_s^2$ corrections is 2.046 GeV.

\setcounter{equation}{0}
\setcounter{section}{2}
\section*{Appendix B. The invariant amplitude $G^{(\pm)}$ of $W^\pm$ decay}
     In this Appendix, we will present the expression of the invariant
amplitude $G^{(+)}$ for $W^+\to f_1\bar f_2$, 
where $f_1$ and $f_2$ have charge $Q_1$ and $Q_2= Q_1-1$, respectively.
The invariant amplitude $G^{(-)}$ for $W^-\to f_4\bar f_3$ is obtained
by replacing  the indices $1\to 3$ and $2\to 4$ in $G^{(+)}$.
The invariant amplitude consists of the Born term, the contribution of
the virtual corrections , the counterterm contribution and
the soft-photon bremsstrahlung contribution.  Decomposing it as, 
\bq
    G^{(+)}(p_W,p_1,p_2) = 1 + G^{(+,{\rm virt})} + G^{(+,{\rm ct})}
                             + G^{(+,{\rm soft})} \label{B.1}
\eq

we find after some calculation, 
\bqa
   G^{(+,{\rm virt})} 
      &=&+{{e^2}\over{16\pi^2}} [Q_1^2\log{{M_W^2}\over{m_1^2}} 
         + Q_2^2\log{{M_W^2}\over{m_2^2}} ]{{\log\lambda^2}\over{M_W^2}} 
           \nonumber \\
      & &+{{e^2}\over{16\pi^2}}   \lbrack
          {{Q_1^2}\over 2} \log^2({{M_W^2}\over{m_1^2}})  
         +{{Q_2^2}\over 2} \log^2({{M_W^2}\over{m_2^2}}) 
         +2Q_1^2 \log{{M_W^2}\over{m_1^2}}+2Q_2^2 \log{{M_W^2}\over{m_2^2}}
           \rbrack \nonumber \\
      & &+{{e^2}\over{16\pi^2}} \lbrack{4\over 3}\pi^2 Q_1Q_2 + 3 
         +{{\ell_1\ell_2}\over{c_W^2s_W^2}}
                [  2(2+{1\over{c_W^2}})(\log(c_W^2)-1) \nonumber\\
      & & ~~~~~~~~~~~~~~~~~  -2(1+{1\over{c_W^2}})^2
		   \{Sp(-c_W^2)+\log(1+c_W^2)\log(c_W^2)\}], \label{B.2}\\
      & &~~~~~~~~~+ {{c_W^2}\over{s_W^2}}[  -2(c_W^2+2)J[1] 
               + 5 + {1\over{c_W^2}}+(2+{1\over{c_W^2}})\log c_W^2 \nonumber\\
      & &~~~~~~~~~~~~~~~~~~  +(2+{1\over{c_W^2}})\int_0^1dx 
                \log(x^2+{{1-x}\over{c_W^2}})]~  \rbrack, \nonumber 
\eqa
where a tiny photon mass, $\lambda$ is indroduced in order to regularize
the infrared singularity.  The left-handed  couplings $\ell_1$ and $\ell_2$
are given by
\bq
    \ell_1 = {1\over 2}-s_W^2Q_1, ~~~~~\ell_2 = -{1\over 2}-s_W^2Q_2,
\label{B.3}
\eq
and $J[1]$ is given by
\bq
   J[1] = \int_0^1{{dx}\over{1-c_W^2x}}
           [-\log x+\log(x^2+{{1-x}\over{c_W^2}})].\label{B.4}
\eq

\bqa
     G^{(+,{\rm ct})}&=&- {{e^2}\over{16\pi^2}} [Q_1^2+Q_2^2+1]
          {{\log\lambda^2}\over{M_W^2}} 
            - {{e^2}\over{16\pi^2}}
            [{3\over 2}Q_1^2\log{{M_W^2}\over{m_1^2}}
            +{3\over 2}Q_2^2\log{{M_W^2}\over{m_2^2}} ]  \nonumber \\
    &&+ {1\over 2}\Delta \alpha(M_W^2) 
         + {{e^2}\over{16\pi^2}}   \lbrack -{1\over 3}-2Q_1^2-2Q_2^2
       +{{\ell_1^2 + \ell_2^2}\over{4s_W^2c_W^2}}+{1\over{4s_W^2}}\rbrack
       \label{B.6}  \\
    && +{1\over 2}\delta Z_{W~f} -{{c_W^2}\over{2s_W^2}}   
      ({{\delta M_{Zf}^2}\over{M_Z^2}}-{{\delta M_{Wf}^2}\over{M_W^2}}),
      \nonumber
\eqa
where the explicit expression of $\delta Z_{W~f}$ and  
${{\delta M_{Zf}^2}\over{M_Z^2}}-{{\delta M_{Wf}^2}\over{M_W^2}}$ can 
be found in Ref.\cite{Fujimoto}. 
\bqa
   G^{(+,{\rm soft})}=-{{e^2}\over{16\pi^2}} 
         & &[\{1+Q_1^2+Q_2^2 -Q_1^2\log{{M_W^2}\over{m_1^2}} 
             -Q_2^2\log{{M_W^2}\over{m_2^2}} \} 
             \log{{4(\Delta E)^2}\over {\lambda^2}} \nonumber\\
         & &+{{Q_1^2}\over 2} \log^2({{4p_{10}^2}\over{m_1^2}}) 
             +{{Q_2^2}\over 2} \log^2({{4p_{20}^2}\over{m_2^2}}) 
             -Q_1^2 \log{{4p_{10}^2}\over{m_1^2}} 
             -Q_2^2 \log{{4p_{20}^2}\over{m_2^2}} \nonumber\\
         & & +(Q_1^2+Q_2^2){{\pi^2}\over 3} 
             + 2(1+Q_1Q_2) Sp(1-{{4E_1E_2}\over{M_W^2}}) \label{B.7}\\
         & & -{1\over\beta} \log({{1+\beta}\over{1-\beta}} )
             - 2\log^2((1+\beta){E\over{M_W}}) \nonumber\\
         & & -2Q_1Sp(1-{2\over{1+\beta}}~{{E_2}\over E} )
             -2Q_1Sp(1-{2\over{1-\beta}}~{{E_2}\over E} )\nonumber \\
         & & +2Q_2Sp(1-{2\over{1+\beta}}~{{E_1}\over E} )
             +2Q_2Sp(1-{2\over{1-\beta}}~{{E_1}\over E} )~],\nonumber
\eqa
where $E_i$ and $\beta$ are defined by (\ref{A.6}) and (\ref{A.2}) 
by setting $k^2_\pm = M_W^2$.
In (\ref{B.2}), (\ref{B.6}) and (\ref{B.7}), the ultraviolet diverging 
parts which cancel from the sum are already discarded.\par
    Removing the dominant fermionic contribution from the sum of 
(\ref{B.2}), (\ref{B.6}) and (\ref{B.7}), one obtains  $G^{(\pm,rest)}$.
It is again decomposed in three parts, which are given as follows.
\bq
   G^{(+,rest)} =  {{e^2}\over{16\pi^2}}(C^{(+,virt)} +C^{(+,ct)}
                                        +C^{(+,soft)}), \label{B.8}
\eq
where
\bqa
    C^{(+,virt)}& =&+{4\over 3}\pi^2 Q_1Q_2 + 3 \nonumber \\
      & &-{{(1-2s^2_WQ_1)(1+2s_W^2Q_2)}\over{4c_W^2s_W^2}}
           [+2(2+{1\over{c_W^2}})(\log(-c_W^2)-1) \nonumber\\
      & &~~~~~~~~~~ -2(1+{1\over{c_W^2}})^2
	             \{ Sp(-c_W^2)+\log(1+c_W^2)\log(c_W^2)\}] \nonumber\\
      & &+ {{c_W^2}\over{s_W^2}}[  -2(c_W^2+2)J[1] 
         + 5 + {1\over{c_W^2}}
           +(2+{1\over{c_W^2}})\log c_W^2 \nonumber\\ 
      & &~~~~~~~~  +(2+{1\over{c_W^2}})\int_0^1dx 
              \log(x^2+{1\over{c_W^2}}(1-x))~]  \nonumber\\
      & =&\cases{ -2.1348 & for~leptonic~ decays \cr
                  -5.0730 & for~hadronic~ decays}  \label{B.9}\\
    C^{(+,ct)}&=& -{1\over 3} -2Q_1^2 -2Q_2^2
             +{{1-2s_W^2+2s_W^4(Q_1^2+Q_2^2)}\over{8s_W^2c_W^2}}
             +{1\over{4s_W^2}}\nonumber \\
   & & +{{16\pi^2}\over{e^2}}[
        {1\over 2}\delta Z_{W~f} -{{c_W^2}\over{2s_W^2}}   
      ({{\delta M_{Zf}^2}\over{M_Z^2}}-{{\delta M_{Wf}^2}\over{M_W^2}})]
       + \log{{\lambda^2}\over{M_W^2}} +{3\over 8}{{m_t^2}\over{s_W^4M_Z^2}},
          \label{B.10}\\
    C^{(+,soft)}& =& -{1\over2}Q_1^2\log^2({{4E_1^2}\over{M_W^2}})
                   +Q_1^2\log({{4E_1^2}\over{M_W^2}})
                   -{1\over2}Q_2^2\log^2({{4E_2^2}\over{M_W^2}})
                   +Q_2^2\log({{4E_2^2}\over{M_W^2}}) \nonumber\\
    & & -(Q_1^2+Q_2^2){{\pi^2}\over 3} 
        - 2(1+Q_1Q_2) Sp(1-{{4E_1E_2}\over{M_W^2}}) \nonumber\\
    & & +{1\over\beta} \log({{1+\beta}\over{1-\beta}} )
        + 2\log^2((1+\beta){E\over{M_W}}) \nonumber\\
    & &+2Q_1Sp(1-{2\over{1+\beta}}~{{E_2}\over E} )
       +2Q_1Sp(1-{2\over{1-\beta}}~{{E_2}\over E} )\nonumber \\
    & &-2Q_2Sp(1-{2\over{1+\beta}}~{{E_1}\over E} )
       -2Q_2Sp(1-{2\over{1-\beta}}~{{E_1}\over E} )~].\label{B.11}
\eqa
The last two terms in the second line of (\ref{B.10}) remove the infrared 
singularity and the dominant top-quark  mass effect present 
in the renormalization constants, making $C^{(+,ct)}$ almost constant
up to the Higgs-mass dependence.


\begin{thebibliography}{99}
\bibitem{Kuschi} M. Kuroda and D. Schildknecht, Nucl. Phys. B531 (1998) 24.
\bibitem{Didi} S. Dittmaier, M. B{\"o}hm and A. Denner, Nucl. Phys. B376
(1992)29; B391 (1993) 483 (E);\hfill\break
J. Fleischer, J.L. Kneur, K. Ko\l odziej, M. Kuroda and D. Schildknecht;
Nucl. Phys. B378 (1992) 443; B426 (1994) 246 (E).
\bibitem{Been} W. Beenakker, A. Denner, S. Dittmaier, R. Mertig and
T. Sack, Nucl. Phys. B410 (1993) 245;\hfill\break
W. Beenakker, A. Denner, S. Dittmaier and R. Mertig, Phys. Lett. B317 (1993)
622.
\bibitem{Berends} W. Beenakker and F.A. Berends, in Physics at LEP2, ed. 
G. Altarelli et al. (CERN 96-01, Geneva 1996), Vol. I, p. 79,
hep-ph/9602351.
\bibitem{Ditt} S. Dittmaier, hep-ph/9811434. 
\bibitem{Ditt1} A. Denner, S. Dittmaier, M. Roth, D. Wackeroth, hep-ph/9904472.
\bibitem{W.Been} W. Beenakker, F.A. Berends and A.P. Chapovski, DTP/98/90,
hep-ph/9811481.
\bibitem{Denner} A. Denner, S. Dittmaier, M. Roth, Nucl. Phys.B519 (1998)
39,
Phys. Lett. B429 (1998) 145.
\bibitem{Fadin} V.S. Fadin and V.A. Khoze, Sov. J. Nucl.Phys. 48 (1988)
309;\hfill\break
V.S. Fadin, V.A. Khoze and A.D. Martin, Phys. Lett B311 (1993)
311; \hfill\break
V.S. Fadin et al., Phys. Rev. D52( 1995) 1377.
\bibitem{Grace} T. Ishikawa et al., GRACE manual vers. 1.0, 1993.
\bibitem{Boehm} M. B\"ohm, A. Denner, T. Sack, W. Beenakker,
 F.A. Berends and H. Knijf,  Nucl. Phys. B304 (1988) 463.
\bibitem{Fleischer} J. Fleischer, F. Jegerlehner and M. Zralek.
Z. Phys. C42 (1989) 409.
\bibitem{Kuroda} M. Kuroda, Analytic expression of the radiative
corrections to the process $e^+e^- \to W^+W^-$ in one-loop order, 1994,
unpublished.  On request, the computer program is available 
from the author.
\bibitem{Ditt2} S. Dittmaier, Acta Phys.Pol. B28(1997) 619.
\bibitem{Denner2} A. Denner, Fortsch. Phys. 41 (1993) 307.
\bibitem{FJZ}  J. Fleischer, F. Jegerlehner and M. Zralek, Z. Phys. 
C42( 1989) 409.
\bibitem{Fujimoto} J. Fujimoto et al., Prog. Theor. Phys. Suppl. 100
(1990) 1.
\bibitem{Jeger} F. Jegerlehner and K. Ko\l odziej, TP-USI/99/01,
hep-ph/9907229. 
\bibitem{BASES}  S. Kawabata, Comput.Phys.Commun. 41 (1986) 127, 
{\it ibid} 88 (1995) 309.
\end{thebibliography}
\end{document}